\documentclass[notitlepage,12pt,aps,pre,showpacs,superscriptaddress,groupedaddress,amsmath,amssymb]{revtex4-1}
\usepackage{epsfig}
\usepackage{graphics}
\usepackage{lipsum}
\usepackage{ulem}
\usepackage{sidecap}
\usepackage{graphicx} 
\usepackage{dcolumn} 
\usepackage{bm}
\usepackage{longtable}
\usepackage{xcolor,colortbl}
\usepackage{color, colortbl}
\definecolor{darkgray}{rgb}{0.66, 0.66, 0.66}
\definecolor{yellow-green}{rgb}{0.6, 0.8, 0.2}
\definecolor{deeppink}{rgb}{1.0, 0.08, 0.58}
\definecolor{darkviolet}{rgb}{0.58, 0.0, 0.83}
\definecolor{darkcyan}{rgb}{0.0, 0.55, 0.55}

\begin{document}
\title{Dynamical modelling and analysis of COVID-19 in India}
\author{R. Gopal$^{1}$, V. K. Chandrasekar$^{1}$ and M. Lakshmanan$^{2}$}
\address{$^1$Centre for Nonlinear Science \& Engineering, School of Electrical \& Electronics Engineering, SASTRA Deemed University, Thanjavur -613 401, Tamilnadu, India.\\$^2$ Department of Nonlinear Dynamics, School of Physics, Bharathidasan University, Tiruchirappalli -620 014, Tamil Nadu, India.\\}
\date{\today}
\begin{abstract}
\par We consider the pandemic spreading of COVID-19 in India after the outbreak of the coronavirus in Wuhan city, China. We estimate the transmission rate of the initial infecting individuals of COVID-19 in India by using the officially reported data at the early stage of the epidemic with the help of Susceptible (S), Exposed (E), Infected (I), and Removed (R)  population model, the so called SEIR dynamical model. Numerical analysis and model verification are performed to calibrate the system parameters with official public information about the number of people infected, and then to evaluate several COVID -19 scenarios potentially applicable to India. Our findings provide an estimation  of the number of infected individuals in the pandemic period of time-line, and also demonstrate the importance of governmental and individual efforts to control the effects and time of the pandemic-related critical situations. We also give special emphasis to individual reactions in the containment process.
\end{abstract}

\pacs{epidemic mathematical model,COVID-19, Governmental action}

\maketitle

\section{Introduction}
COVID-19, a disease caused by the severe acute respiratory syndrome coronavirus 2 (SARS-COV 2), was first identified in December 2019 in Wuhan, the capital of Hubei, China, and has since spread globally~\cite{cohen}. The World Health Organization (WHO) announced COVID-19 as an international public health emergency on $30^{th}$ January, 2020, and subsequently a pandemic on $11^{th}$ March, 2020~\cite{who}. The number of patients is growing exponentially and thousands of people are losing their lives in many a countries, almost every day due to COVID-19~\cite{lin,li,fer}.  The complexity of the situation can be realized from the fact that as on $15^{th}$ May, 2020 the number of coronavirus cases worldwide has reached a staggering~ 46,39,427, with infected patients being~25,64,442 and more than~3,08,810 confirmed deaths have been reported  due to this disease~\cite{pop}. Moreover, the outbreak has also spread to more than two hundred countries~\cite{pop} (Note added in revision: In continuation, the number of coronavirus cases reached as 46,403,652 with currently infected patients being 12,079,415 and 1,211,421 confirmed death cases as on 2$^{nd}$ November, 2020).

Much work indicates that COVID-19 could spread from animal to human (zoonotic)~\cite{rothan}. In addition, a rapid increase of COVID-19 infections show the main finding that secondary transmission can occur through human-to-human contacts or through droplets transmitted by coughing or sneezing from an infected person~\cite{rothan,li} or even when an infected person speaks to a non-infected one. With the above trend, this spread of human-to-human disease is growing significantly almost everywhere in the world and the infection is rapidly increasing in many countries through local transmission~\cite{li}.

In India, the infection due to COVID-19 was first identified on $02^{nd}$ March, 2020. The Indian Government announced a 21-day country-wide lock-down as a preventive measure for the COVID-19 outbreak on $24^{th}$ March, 2020. The aim of the lockdown was to slow down the spread of the novel coronavirus, to allow the Government to follow a multi-pronged strategy to add more beds in its hospital network, to increase the development of the COVID-19 test kits and of personal protective equipments (PPEs) for health workers, etc. The government frequently uses different platforms to keep the public aware of COVID -19. In India, conditions including very high population density in urban areas,  unavailability of vaccines and inadequate data about the disease's transmitting process also make it a herculean task to adequately fight the disease.

Mathematical simulations were often used to forecast the effects of various epidemics and also to test the efficacy of the different prevention approaches in reducing the burden of the epidemics~\cite{ker}. Recently, concerning the current COVID-19 pandemic, considerable research has been carried out using actual occurrence from the impacted countries, analyzing various aspects of the epidemic, as well as assessing the impact of preventive approaches adopted in order to limit the epidemic in the countries concerned~\cite{arxiv}. In particular, various kinds of dynamical models have been employed, essentially considering nonlinear governing equations. For instance, the nature of human coronavirus infection and defining contact between human cells and the virus have been described in Ref~\cite{rihan}. The statistical model for estimating virus transmission, taking into account a condensed version of the bats-hosts-reservoir-people transmission model, known as a reservoir-people model was also reported in Ref~\cite{chen}.

Susceptible-exposed-infectious-recovered (SEIR) model is an important tool for tackling coronavirus transmission statistical simulations. Lin and his co-workers have proposed a method to develop analytical models for the COVID-19 outbreak in Wuhan taking into account human behavioral responses and government decisions, such as holiday extension, travel restriction, hospitalization, and quarantine~\cite{lin}. Some  recent  studies based on these works  use numerical simulations   and attempt to  provide  a reliable  real-time  forecast  of  COVID-19 cases in various countries with the help of this mathematical model. For instance, Savi \emph{et al} have obtained the general transmission of the novel coronavirus to test various scenarios of coronavirus propagation in different countries, taking into account the model testing of the evolution of infected populations in China, Italy, Iran and Brazil based on government and individual reactions~\cite{savi}.

In this work, we consider the situation in India starting from the initial outbreak period and fitted SEIR model to the daily infected cases reported between $2^{nd}$ March, 2020 to  $15^{th}$ May, 2020. We estimate the basic transmission rate of COVID-19 in the initial stage of the epidemic  between $2^{nd}$ March, 2020 and $24^{th}$ March, 2020. Further, the general propagation of the novel coronavirus is also studied in our investigation to evaluate different scenarios of the propagation of coronavirus. In addition, the model verification takes into consideration the evolution of the infected population and simulates different scenarios based on the rate of transmission and the governmental and, especially individual reactions. Finally it proposes potential evolution of the spread and possible mitigations, specifically emphasizing the significance of individual reactions at the societal level during the pandemic period which includes both lock-down and various unlock periods.

The structure of the paper is as follows. In section II, we briefly describe the  SEIR dynamical model. The estimation of transmission rate and detailed numerical analysis of the mathematical model with actual data is described in section III. Finally, we discuss the findings obtained from our study in Section IV. In the Appendix (added in the revision), we also provide details of our study on the number of  infected individuals during the lock-down and unlock periods starting from March, $25^{th}$, 2020 to October 31 2020 and further up to December $27^{th}$, 2020.

\begin{table}
\centering
\caption{Summary table of the parameters discussed in Eqs.~(\ref{eq1}) and (\ref{eq2})}
\label{table1}
\begin{tabular}{ccc}
\toprule
Parameter & Description & value/remarks/reference \\
\toprule 
$N_{0}$ & Initial number of population & India populations~\cite{pop1}\\

$S_{0}$ & Initial number of susceptible population & $0.9N_{0}$ (constant)  \\

$E_{0}$ & Exposed persons for each infected person & $20I_{0}$ (assumed) \\
 
$I_{0}$ &Initial state of infected persons & 3~\cite{pop1} \\
 
$\alpha$ & Government action strength  & varied in each lock-down period\\

k & intensity of individual reaction & 1117.3~\cite{lin,savi} \\

$\sigma^{-1}$ & Mean latent period & 3 (days) \\
 
$\gamma^{-1}$ & Mean infectious period & 5 (days) \\ 
 
$d$ & Proportion of severe cases & 0.2  \\

$\lambda^{-1}$ & Mean duration of public reaction & 11.2 (days)  \\

\toprule
\end{tabular}
\end{table}

\begin{figure*}
\centering
\includegraphics[width=0.7\columnwidth]{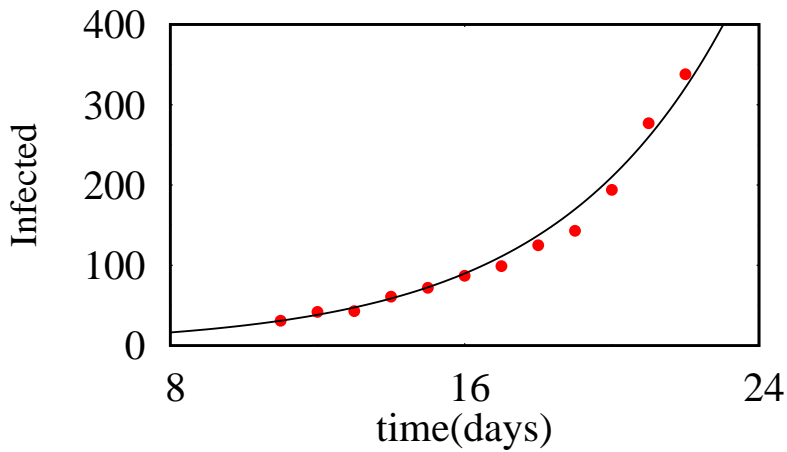}
\caption{The number of infected individuals (excluding number of initial quarantined people) after the first few days of outbreak on $2^{nd}$ March, 2020.~ Data were fitted by the nonlinear least-square fit for India. The solid black line is fitting data and red circles denote real data of the number of infected individuals.}
\label{fig1}
\end{figure*}

\section{The Dynamical model}

Lin and his co-workers have proposed a susceptible-exposed-infectious-removed (SEIR) model to explain coronavirus disease (COVID-19)~\cite{lin}. This model was inspired by He \emph{et al's} original influenza model~\cite{he,he1}.  The SEIR model also has two supplementary terms: $D$ is a public perception of risk with respect to serious cases and deaths, and $C$ is the number of recorded and unreported incidents. In addition,  $S$ is the susceptible population, $E$ is the population exposed, $I$ is the currently infectious population (excluding the recovered and death cases)  and $R$ is the population removed which includes both the cases of recovered and deaths. The simplified version of the governing equations takes account of the interaction between all these populations, and is represented by the following set of coupled nonlinear differential equations~\cite{lin},

\begin{subequations}
\begin{eqnarray}
&&\dot{S}=-\beta(t)\frac{SI}{N},  \\
\label{eq1b}
&&\dot{E}=\beta(t)\frac{SI}{N}-\sigma~E ,  \\
\label{eq1c}
&&\dot{I}=\sigma E-\gamma I , \\
&&\dot{R}=\gamma_{R} I,  \\
&&\dot{D}=d\gamma I-\lambda D ,  \\
&&\dot{C}=\sigma E,  
\label{eq1}
\end{eqnarray}
\label{eq1}
\end{subequations}
where $\gamma$ is the mean infectious period, $\gamma_{R}$ is the delayed removed period, which denotes the relation between removed population and the infected one, $\sigma$ is the mean latent period, $d$ denotes the proportion of severe cases and $\lambda$ is the mean duration of public reaction~\cite{li,lin,savi}.

In Eq.(\ref{eq1}), $\beta(t)$ denotes the transmission rate function which incorporates the impact of governmental action $(1-\alpha)$, and the individual action, which is denoted by the function $\left (1-\frac{D}{N} \right )  ^{k}$~\cite{lin,savi}.   Here, the parameter $k$ defines the intensity of individual reaction, which is measured on a scale of 0 to $10^5$ with a normal value of 1117.3 obtained from previous and recent epidemic and pandemic studies~\cite{he,lin}.  We also assume that the effect of governmental action is different during different  lock-down periods. Therefore, the transmission rate $\beta(t)$ is defined as

\begin{align}
\beta(t)=\beta_{0}(1-\alpha) \left (1-\frac{D}{N}\right)^{k}. 
\label{eq2}
\end{align} 
The value of $\beta_{0}$ is derived by assuming that the basic reproduction number is $R_{0}=\frac{\beta_{0}}{\gamma}$, which measures the average number of new infections generated by each infected person. The values of the system parameters are mentioned in Table~\ref{table1}, based on the information deduced from the references in~\cite{he,li,lin,savi,das}.

The above parameters must be modified for each state/country which is important for the analysis of COVID-19. In general, the physical meanings of the parameters are based on a variety of facts, identifying which constitute a difficult task~\cite{lin}. It should be also pointed out in this regard that the real data has spatial aspects that are not covered by the above set of governing equations. Consequently, this kind of study is a sort of average activity that needs careful adjustment to suit real data as followed by earlier studies~\cite{lin,savi}.

In our present study, we use step-like functions to define certain parameters that allow for a proper representation of different scenarios, especially the rate of transmission. It is also important to remember that all governmental or individual decisions have a delayed impact on the dynamics of the system. Further, virus mutation is another important factor related to the definition of COVID-19 dynamics which can affect the reaction of the system significantly but is not discussed in our present study~\cite{savi}. The next sections treat  the COVID-19 dynamics considering two different objectives. To start with, we identify the initial value of transmission rate with the data from initially infected people, and then it examines various scenarios for the COVID -19 situation in India, using different transmission rates and government and individual action strengths.

\begin{figure*}
\centering
\includegraphics[width=0.7\columnwidth]{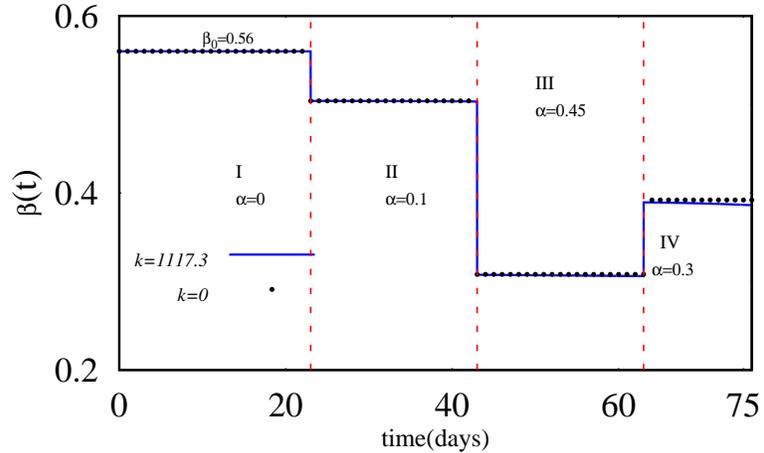}
\caption{Variation of transmission rate considered in our model (\ref{eq1}), through time with the impact of  $\alpha$ in (\ref{eq2}) with respect to initial transmission value $\beta_{0}=0.56$. The continuous curve is for intensity of individual reaction $k=1117.3$ and the dotted curve is for $k=0$.} 
\label{fig2}
\end{figure*}

\begin{figure*}
\centering
\includegraphics[width=1.0\columnwidth]{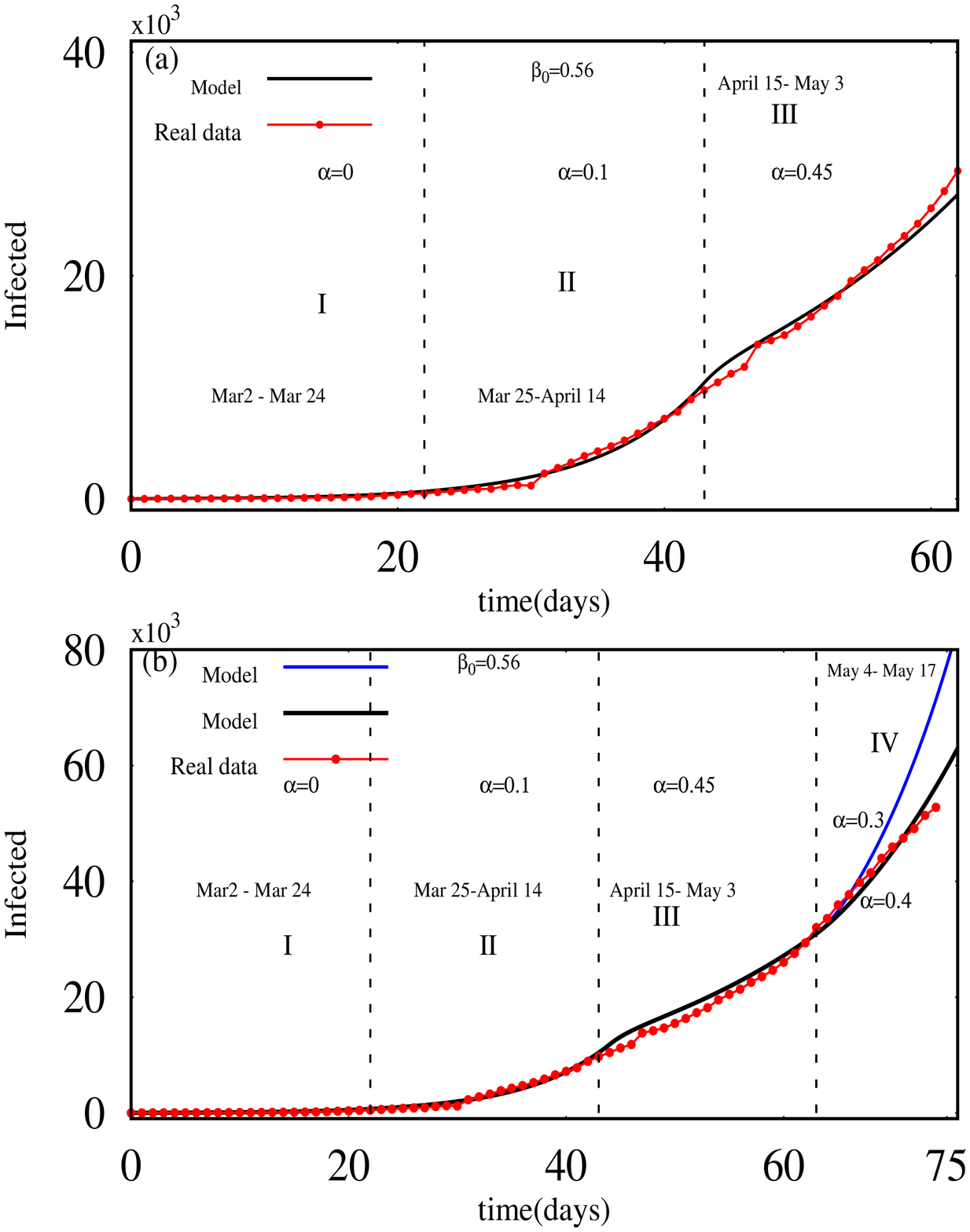}
\caption{Numerical simulation of the number of infected individual people (excluding both recovered and deaths) with the value of initial transmission rate taken as $\beta_{0}=0.56$ and for different governmental action strengths before and after the lock-down period. The prediction of the mathematical model (continuous blue and black curve) compared with actual data (red dotted curve) for the daily infected people in India upto (a) $3^{rd}$ May, 2020 and, (b)  $17^{th}$ May, 2020 is depicted. The strength of individual reaction is considered as $k = 1117.3$ and  the remaining parameters are taken from Table \ref{table1}.}
\label{fig3}
\end{figure*}

\section{Data analysis with numerical model} 
As a first phase of the established study,  model verification is performed using information available on India covid-19 tracker and worldometers.info~\cite{pop}. We follow the SEIR implementation methodology described for various countries in~\cite{savi}. The model employed for simulations is with a total population of India at $N\approx 139 \times 10^{7}$~\cite{pop1}, and 3 infected COVID-19 confirmed as on 02-Mar-2020. The initial state is taken to be with $I_{0}=3$ and a susceptible initial population is assumed to be $S_{0}=0.9N$. Another information needed for the model is the number of individuals exposed to each infected person. Each infected individual is believed to  have the potential to infect a further 20 individuals, $E_{0}=20I_{0}$~\cite{lin,savi}. Initially, we started with the estimation of the initial transmission rate.

\subsection{Estimation of initial transmission rate}

To start with, we note that, intensified precautionary measures to curb the spread of COVID -19 was carried out in India. In particular, when patients were found at a location, the Committee of Physicians and Experts checked those people who had contacts with the patients. After that, both the patients and the contaminated people were quarantined in a hospital or in other isolated areas. In India's data published, in the initial few days, the active cases were automatically moved to quarantined cases~\cite{gov}. Therefore, the active cases correlate after a few days with the omitted quarantined cases. From the recorded data we can match real infected individuals (excluding quarantined people) as a function of time at the early stage of the disease spreading bar the initial quarantined people~\cite{chae}.

In order to solve the model (\ref{eq1}), we consider the susceptible population to be close to the overall population in the early stages of transmission of the disease~\cite{chae}, and we can rewrite the dynamic equation of the epidemic  with $\frac{S}{N}\approx 1$.

In this case, from Eqs.~(\ref{eq1c}) and (\ref{eq1b}), the equation for the presently infectious population (which excludes the recovered and death cases) becomes the second order linear ordinary differential equation of damped linear type,
\begin{eqnarray}
\ddot{I}+(\gamma+\sigma)\dot{I} -\sigma(\beta-\gamma)I=0. \label{eq01}
\end{eqnarray}
By integrating the above equation, we obtain the solution as~\cite{lak}
\begin{eqnarray}
I(t)=I_{01}e^{\frac{-1}{2}(\gamma+\sigma-\sqrt{(\gamma-\sigma)^2+4\sigma\beta})t} +I_{02}e^{\frac{-1}{2}(\gamma+\sigma+\sqrt{(\gamma-\sigma)^2+4\sigma\beta})t},
\label{eq02}
\end{eqnarray}
where $I(0)=I_{01}+ I_{02}$ is the initial number of individual infected people. Now the curve is fitted with the initial real infectious population data available in~\cite{pop}. Further we identify an exact fit with the curve $3 e^{0.473 t}$. From this we obtain the parameter values as $I_{01}=3$, $I_{02}=0$ and the initial value of $\beta$ is  0.56. This $\beta$ value can be considered as the initial transmission rate $\beta_0$ in (\ref{eq2}). It gives the reproduction number $R_0=\frac{\beta_{0}}{\gamma}=2.8$.  One may note that a few recent studies suggest a reproduction rate of $R_{0}$ about 2.52 for the first 22 days before lockdown, and then identify the value of $R_{0}$ to vary between 1.9 to 3.0 in various places in India, due to the variation of transmission rates~\cite{das,liu,chae}, but in our analysis the estimation of $R_{0}$ from the  model is confirmed with real data of daily infected individuals (excluding recovered and death cases). In Fig. \ref{fig1} we represent the curve fit of the infected people as a function of time at the early stage of the disease spreading in India.

\begin{figure*}
\centering
\includegraphics[width=1.0\columnwidth]{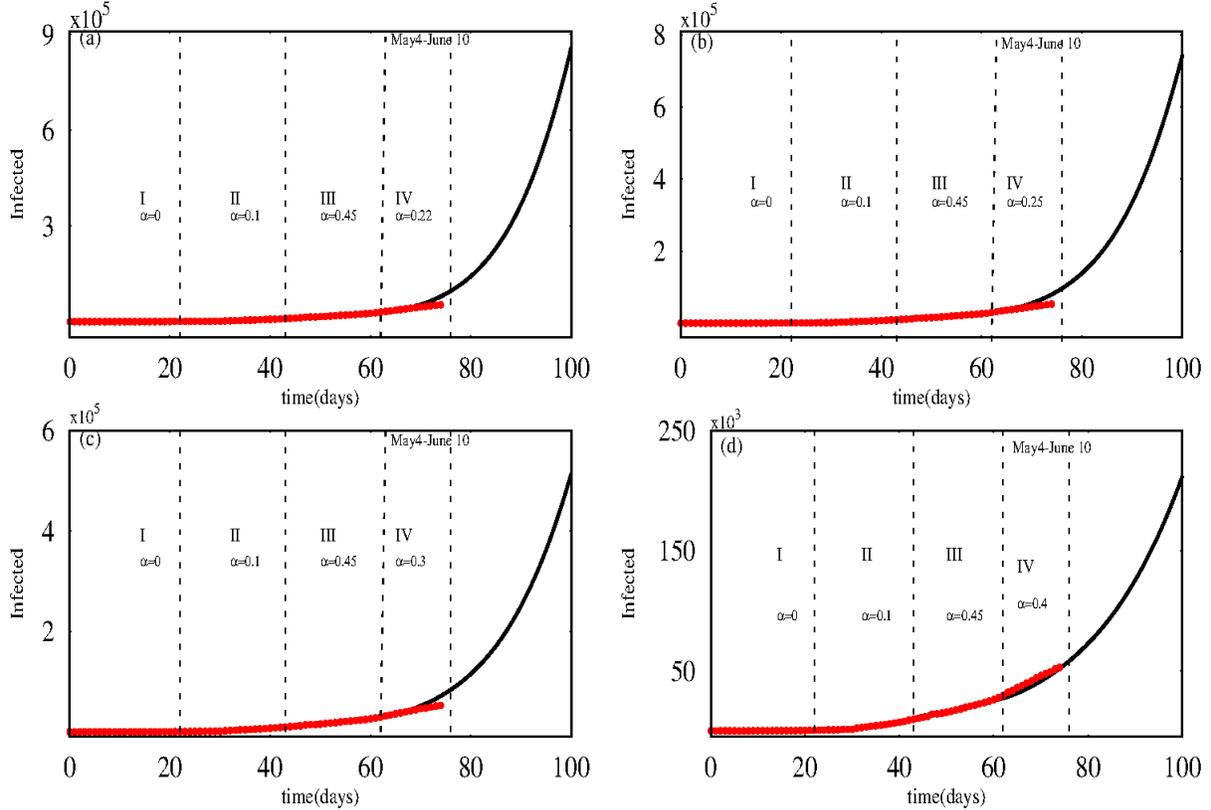}
\caption{Numerical simulation of the evolution of number of infected individuals with the value of initial transmission rate $\beta_{0}=0.56$. The red curve denotes actual data for the daily infected people (excluding recovered and death cases) in India  up-to $15^{th}$ May, 2020. Note that this simulation (See column IV) shows the rate of newly infected population after May 3 with respect to the values of government action strength $\alpha$ in (\ref{eq2}) with (a) $\alpha=0.22$, (b) $\alpha=0.25$, (c) $\alpha=0.3$, and (d) $\alpha=0.4$. The strength of individual reaction considered is a rather low value of k = 1117.3 and the remaining parameters are taken from Table \ref{table1}.}
\label{fig4}
\end{figure*}

\subsection{Verification through Simulations}

In this section, the first scenario for model verification  based on the results of India is presented. It should be noted that our analysis considers all infected cases in entire India, and it is not restricted to any specific state or place. The parameters which are used in the model are specified in Table \ref{table1}, and these parameters should be treated as average since they are appropriate for the country as a whole. Initially, we started with the value of transmission rate starting with $\beta_{0}=0.56$ and the value attributable to governmental action strength is also considered as different  during the periods before and after lock-down. Fig. \ref{fig2} charts the transmission rate $\beta(t)$,   before and after the lock-down period with respect to various governmental action strengths. One may note that one important constraint on the  parameter $\beta(t)$ is that this variable should be a step-like function with time due to the impact of government action strength. Therefore, using the daily COVID-19 incidence data, numerical simulation is carried out for the model (\ref{eq2}) with the value of $\beta_{0}=0.56$. Fig.~\ref{fig3} shows the evolution of the number of infected individuals (excluding both recovered and death population), indicating a strong agreement between simulation and actual data. The infected people regions are marked as I, II,III and IV and they correspond to the periods of Mar 02 -Mar 24 (before lock-down), Mar 25-April 14 (first lock-down), April 15 - May 3 (second lock-down) and May 4 - May 17 (third lock-down), respectively.

\begin{figure*}
\centering
\includegraphics[width=0.7\columnwidth]{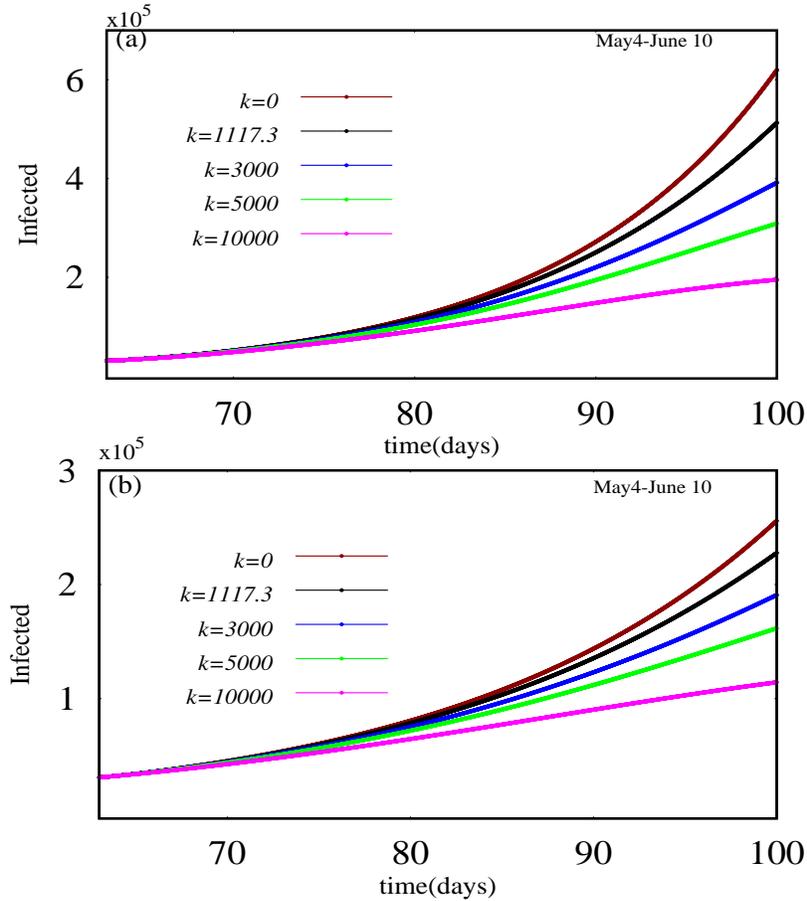}
\caption{Numerical simulation of the evolution of the number of infected individuals with the specific value of initial transmission rate $\beta_{0}=0.56$, after May 3 (expanded view of column IV in Figs.~\ref{fig4}(c)-(d)) with respect to different values of intensity of individual response $k$ in Eq.(\ref{eq2}) for different values of government action strength in (a) $\alpha=0.3$, and (b) $\alpha=0.4$. The remaining parameters are taken from Table \ref{table1}.}
\label{fig5}
\end{figure*}

\begin{figure*}
\centering
\includegraphics[width=0.7\columnwidth]{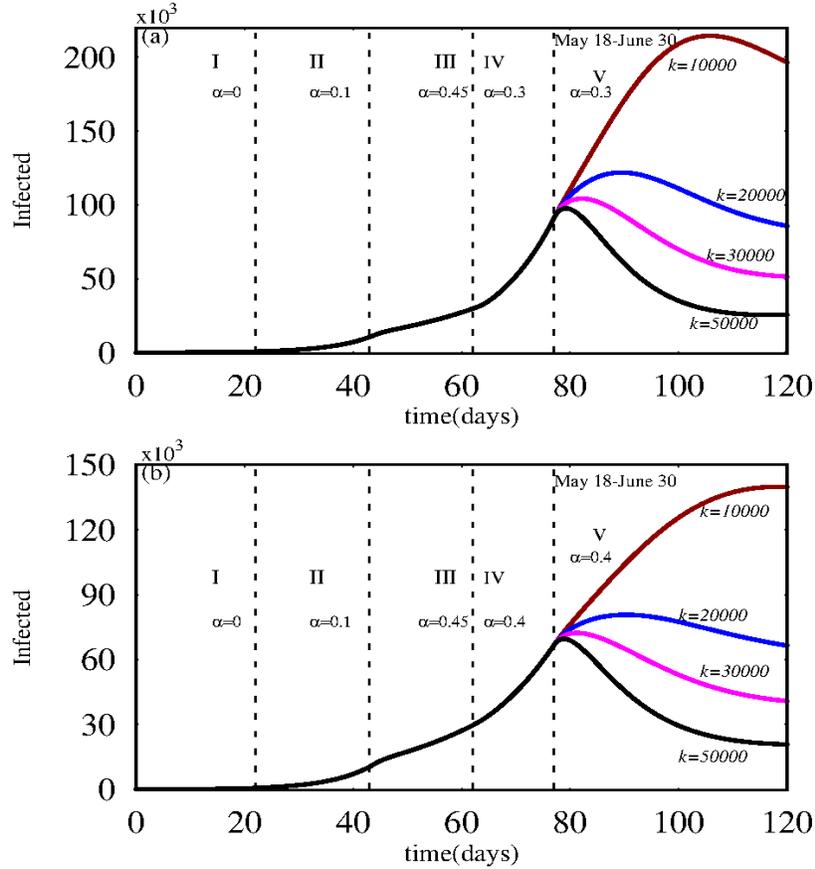}
\caption{Numerical simulation of the evolution of the number of infected individuals for the value of initial transmission rate $\beta_{0}=0.56$ and $k=1117.3$ (I~-~IV). Note that this simulation shows rate of infected people after May 17 (see column V) with respect to the different values  individual reaction in Eq.~(\ref{eq2}) with (a) $\alpha=0.3$ and (b) $\alpha=0.4$. The remaining parameters are taken from Table \ref{table1}.}
\label{fig6}
\end{figure*}

It is apparent that our estimation of the number of infected individuals in the model (1),  with basic transmission rate, system parameters and different governmental action strengths upto  the end of the second lock-down period starting from initial outbreak (Mar 2 -May 3) shows good agreement with the actual data of infected individuals (See Fig.\ref{fig3}(a)).  This study helps to predict newly infected individuals in the near future.  For instance, we choose  low values of governmental action strengths, $\alpha=0.3$ and $0.4$, in the third lock-down period (due to the minimum relaxation given by the government, when certain activities were permitted and restricted in each zone) compared to $\alpha = 0.45$ during the second lock-down period, and our model study predicts the number of infected individuals reasonably well with the real data for $\alpha=0.4$ in the region IV  (See black curve in Fig.\ref{fig3}(b)). Note that in Figs.\ref{fig3}, we have also included the strength of individual reaction at a rather low value of $k = 1117.3$, see Eq. (\ref{eq2}) above, but the contribution due to this term is only minimal.  A sample summary of the actual infected people in India~\cite{pop}, and prediction of infected individuals from the  model (\ref{eq1}) are given in Table.\ref{table2}.

\begin{table}
\centering
\caption{Sample summary of actual data of COVID-19 pandemic in India~\cite{pop} and predicted infected people (I(t)) from SEIR model given by Eqs.~ (\ref{eq1}) and (\ref{eq2})}
\label{table2}
\begin{tabular}{ccc}
\toprule
Date &   Number of Infected People~\cite{pop}(worldometers) &  SEIR Model (I(t)) \\
\toprule 
04/03/2020 & 26 & 29 \\
10/03/2020 & 58 & 84 \\
24/03/2020 & 486 & 679 \\
2/04/2020 & 2280 & 2249 \\
11/04/2020 & 7189 & 7103 \\
14/04/2020 & 9735 & 10417 \\
15/04/2020 & 10440 & 11564 \\
19/04/2020 & 14202 & 14660 \\ 
23/04/2020 & 17306 & 17575 \\
24/04/2020 & 18171 & 18370 \\
25/04/2020 & 19519 & 19198 \\
03/05/2020 & 29339 & 27261 \\
10/05/2020 & 43980 & 43910 \\
13/05/2020 & 49104 & 55958 ($\alpha=0.3$); 46368 ($\alpha=0.4$)\\
14/05/2020 & 51379 & 60668 ($\alpha=0.3$); 49073 ($\alpha=0.4$)\\
15/05/2020 & 52773 & 65767 ($\alpha=0.3$); 51904 ($\alpha=0.4$)\\
\toprule
\end{tabular}
\end{table}

\subsection{Role of governmental action and individual reaction}

In addition, we also analyze the possibility of impact of prevention approaches in reducing new cases infected with COVID-19 via the above mathematical model, after $3^{rd}$ May,  2020 (For instance $4^{th}$ May, 2020 - $10^{th}$ June, 2020, and beyond, marked as region IV in Figs. \ref{fig4}(a)-(d)). Strategies for prevention include preventive mechanisms such as lock-down, information campaign by newspapers and television, adequate hand sanitation, social distancing etc. which results in slowing down the COVID-19 transmission process. These strategies of prevention as modeled in terms of the parameter $\alpha$  and $k$ in Eq.~(\ref{eq2}), which imply that there will be a reduction in the transmission rate $\beta(t)$.  Now, we start with the efficiency of prevention by varying government action strength alone, keeping $k$ at fixed low strength $k=1117.3$~\cite{lin}. For such low values of strengths, that is for $\alpha=0.22$ and $\alpha=0.25$, it is observed that the number of infected individuals will peak around $8\times 10^{5}$ and $7\times 10^{5}$ respectively by June 10 (see Fig.\ref{fig4}(a)-(b)).  If the strength is increased (for $\alpha = 0.30$), the peak of the infected/active people cases  may decrease and the occurrence of the peak is shifted down to $5\times 10^{5}$  on June 10 (see Fig.\ref{fig4}(c)).~Now, we consider a further increased value of government action strength (for $\alpha = 0.4$), and the newly infected cases can decrease and reach around $2.5\times 10^{5}$  by June 10 (see Fig.\ref{fig4}(d)).

Analyzing the results of the above, the present dynamical model clearly shows that when the value of $\alpha$ is  reduced, new infected cases continue to quickly increase, while a larger value of $\alpha$ decreases the infected cases to a considerable extent. However, the later will not also help in reducing the infection to approach zero unless perhaps $\alpha$ approaches a value close to unity. Moreover, it is not practical to keep increasing the governmental action to a higher and higher level even  in the fourth lock-down period and further due to the necessity of opening up the economic front for survival of the nation. On the other hand, with relaxation of governmental action over time, particularly in the fourth lock-down period and subsequent period,  the above type of prevention method alone will not end up in the eradication of the disease.

In the above, in Figs.~\ref{fig4}, we have shown our simulations, based on the various governmental action strengths with fixed low value of intensity of individual reaction ($k=1117.3$) in Eq.(\ref{eq2}). Individual reactions or behavior can include social distancing, personal hygiene, health habits and avoiding crowded places and so on~\cite{who}.  It will also include alerting fellow citizens to wear masks, to follow personal hygiene and social distancing, political and social organizations urging fellow citizens to follow social norms and so on. Now, we also address the importance of individual reactions with low value of governmental action strength $\alpha=0.3$ and $\alpha=0.4$ in the region V (which corresponds to the period beyond May 18) in Figs.~\ref{fig5}. We observe from Figs.~\ref{fig5} that  the number of infected people increases for no action of individual response (for $k=0$),  while the number of infected people decreases when the value of individual response increases to $k$=3000, 5000 and 10000. From these figures we also learn that individual behavioral responses are also very important along with governmental action.

In the above scenario, we further considered different low values of government action strength $\alpha$ in the region IV in Figs.\ref{fig4}(a)-(d) (i-e May 3- June 10) and various values of intensity of individual reaction in the same region in Figs.~\ref{fig5}(a)-(b). The corresponding figures show that the disease continues to infect more and more people due to low value of governmental action strength, but it can become controllable with respect to appropriate individual reactions. In order to break the chain of infection spread and to get more controllable handle on infected individuals one may choose immediate action of the individual reaction response after third lock-down period, that is range IV, and consider appropriate values of $\alpha$ and $k$ in the time window V in Figs.~\ref{fig6}.  For $\alpha=0.3$ or 0.4 and $k=10000$ or $k=20000$  the newly infected cases tend to decrease within a few weeks, after May 18 (see Figs.~\ref{fig6}(a)-(b)).  If the  individual reaction is increased further ($k=30000$ and $k=50000$) we see that the disease can be effectively eradicated within 1 to 2 months from $18^{th}$ May, 2020 (see Figs.~\ref{fig6}(a)-(b)).

 Based on our analysis in Figs.~\ref{fig4}-\ref{fig6}, we find that if we introduce appropriate values of individual reaction strength $k$, then even for low values of governmental action, the reduction can become substantially impressive and one can approach a regime of complete controlling of the disease in a reasonably short period in the absence of appropriate vaccination and so on.

\section{Conclusion}
On $3^{rd}$ May, 2020 the total number of active infected cases registered for COVID-19 and deceased cases in India were 29,339 and 1391, respectively and on May $15^{th}$ they stood at 52,773 and 2753, respectively and on November 04, 2020 they stood at 5,33,787 and 1,23,611, respectively~\cite{gov}. This amount of rise in the active infected cases has happened after some minimal relaxation in the government lock-down, and  several hundred  new cases are reported every day from different locations across India. Our study analyzed the effect of lock-down days on the spread of COVID-19 disease in India. Therefore,  predictive mathematical/dynamical models can also provide useful insights to strengthen our understanding of COVID-19 transmission and control.

In our study, we considered a dynamical model of Susceptible-Exposed-Infectious-Removed (SEIR) spreading epidemic, and estimated the initial rate of COVID-19 transmission by considering the initially infected people in India. In addition, a verification procedure is also performed with respect to different transmission rate values based on the data available from India. Our findings also indicate that the government and in particular individual efforts are important in reducing infected populations and also in reducing the overall epidemic period.  In addition, we would like to note that these kind of epidemic mathematical models and their predictive simulations are also valuable resources which can be helpful for public health planning and in governmental as well as individual acts. Further,  our study also shows that the COVID-19 pandemic can be suppressed by a lock-down.

Our model and current data seem to indicate that the confirmed infected individuals continue to grow in India every day, in spite of rapid response by the government to the pandemic through various quarantine measures, nationwide lock-down and risk-based zoning and so on when the individual reaction rate ($k$) is taken as low. But we find that for appropriate increased values of individual reaction, even with low governmental action strength, there can be dramatic reduction in the total number of infected people. Depending upon the increased individual contribution the disease can be effectively controlled in a rather short period. It is then imperative that society as a whole contributes its might by simple social measures, besides appropriate governmental action. These combined efforts can contribute towards a total control of the disease in a short period in India and perhaps elsewhere as well.

\section*{Acknowledgements}
The work of V.K.C. forms part of a research project sponsored by SERB-DST-MATRICS Grant No. MTR/2018/000676. M.L. wishes to thank the Department of Science and Technology for the award of a SERB Distinguished Fellowship under Grant No.SB/DF/04/2017. 

\begin{figure*}
\centering
\includegraphics[width=1.0\columnwidth]{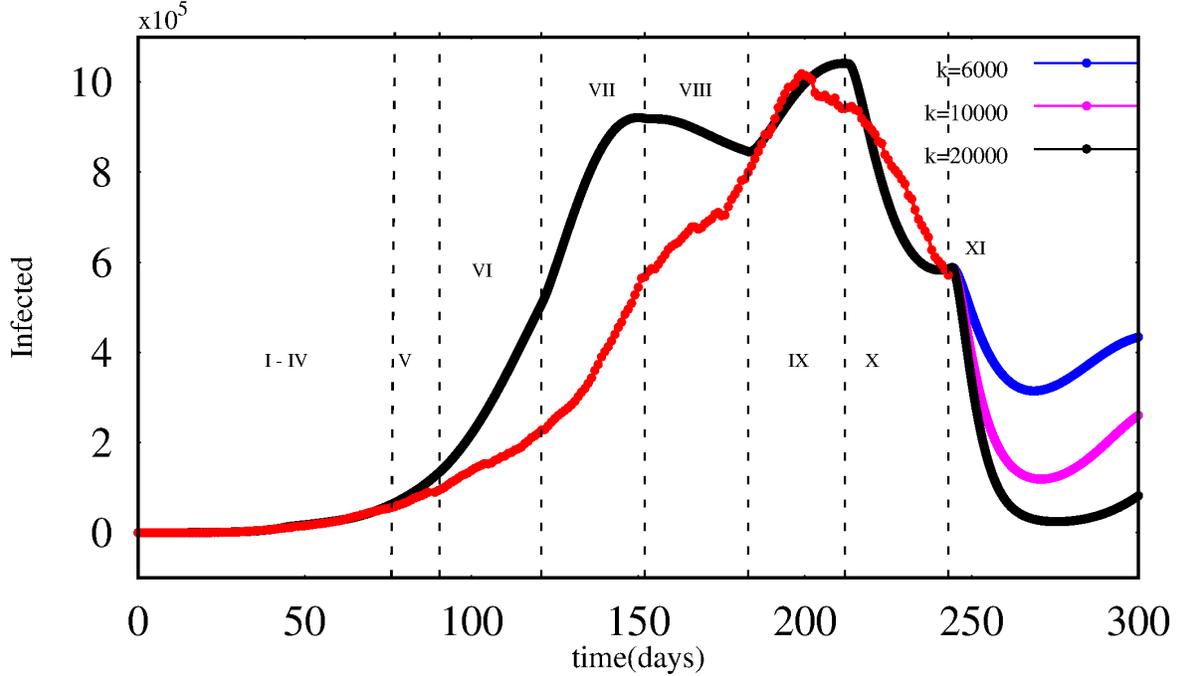}
\caption{Numerical simulation of infected people with the value of initial transmission value $\beta_{0}=0.56$. The red curve denotes actual data from refs~\cite{pop,gov} for the daily infected people in India  up-to Oct 31,2020. Note that the present present simulation (I-XI column) shows the rate of infected people in the timeline, from Mar 02,2020 to Dec 27,2020 with respect to the various values of government action strengths $\alpha$ and individual reaction $k$ as given in Table III. If the strength of the individual reaction $k$ is increased further (after October 31,2020), the number of the infected individuals may decrease and come to an end in a short period of time (black curve in region XI).}
\label{fig7}
\end{figure*}

\section*{Appendix-I (Added in the revision)}
\subsection{Study of infected individuals during lock-down and unlock periods}

In order to get a clear insight on the estimation of our prediction, Fig.~\ref{fig7} shows the rate of infected individual people with respect to timeline of various lock-down and unlock periods. It shows that various intervention strategies (lock-down, spreading awareness program, public reaction, proper hand sanitization, etc), including measures like governmental action strength $\alpha$ and intensity of individual reaction $k$ play an important role in controlling the number of infected individual people in India. 

Staring from the initial date of outbreak, we consider a period of 300 days as timeline  and study the impact of various interventions during this period. Speaking in terms of the actual dates, we consider the time period between $2^{nd}$ March 2020 to $27^{th}$ December 2020.  Our study shows that with comparatively low values of the parameters $\alpha$ and $k$ the time of occurrence of the peak of the outbreak happens during the unlock period Unlock 4.0, and it is represented as region IX in Fig.~\ref{fig7}. If the strength of the intensity of individual reaction  increases further from $k$ = 2500 in region IX to 3500 in region X and then to 6000 and then to 10000 or even 20000 in the region  XI, the number of infected individuals decreases rather quickly and can reach a minimum at the end of December 2020 (The increased value of individual reaction $k$ may be attributed to increased awareness among people and more number of individuals following basic hygiene,etc..). Our results seem to agree well with a recent COVID-19 data on India~\cite{who,gov}. Further, our study also shows that when the value of $k$ is decreased further in the forthcoming period in the XI window region (Fig.\ref{fig7}), the new cases tend to increase and it will show the possibility of increased occurrence of  infected individuals rapidly (blue and pink curves in region XI).

\setlength{\tabcolsep}{10pt} 
\renewcommand{\arraystretch}{1.5}
\begin{table}
\centering
\caption{Values of governmental action strength $\alpha$  and, intensity of individual reaction $k$ used during No lock-down (No LD), lock-down (LD) and unlock period (UL) in Eq.~(\ref{eq2}).}
\label{table3}
\begin{tabular}{|c|c|c|c|c|c|}
\hline
Phases & Timeline & Figure & Region  & value of $\alpha$ & value of $k$\\ [2ex]
\hline 
No LD & 02 March - 24 March & \ref{fig3}(b) & I & 0.0 & 1117.3 \\
LD-1 & 25 March - 14 April & \ref{fig3}(b)  & II & 0.1 & 1117.3 \\
LD-2 & 15 April - 3 May    & \ref{fig3}(b) & III & 0.45 & 1117.3\\
LD-3 & 4-17 May           & \ref{fig3}(b) & IV & 0.40 & 1117.3\\
LD-4 & 18-31 May         & \ref{fig7} & V   & 0.40 & 2000\\
UL-1 & 1-30 June         & \ref{fig7} & VI  & 0.38 & 2100\\
UL-2 & 1-31 July         & \ref{fig7} & VII & 0.30 & 2200\\
UL-3 & 1-31 August       & \ref{fig7} & VIII & 0.25 & 2300\\ 
UL-4 & 1-30 September    & \ref{fig7} & IX  & 0.11 & 2500\\
UL-5 & 1-31 October      & \ref{fig7} & X   & 0.10 & 3500\\
UL-6 & 1 November-       & \ref{fig7} & XI  & 0.10 &  6000  (blue)  \\
     &                   & \ref{fig7} & XI &  0.10 &  10000  (pink) \\
     &                   & \ref{fig7} & XI &  0.10 &  20000  (black) \\[5ex]
\hline
\end{tabular}
\end{table}

\end{document}